# Authentication Using Piggy Bank Approach to Secure Double-Lock Cryptography

Subhash Kak

**Abstract.** The piggy bank idea allows one-way encryption of information that can be accessed only by authorized parties. Here we show how the piggy bank idea can be used to authenticate parties to counter man-in-the-middle (MIM) attack that can jeopardize the double-lock cryptography protocol. We call this method double-signature double lock cryptography and it can be implemented in ways that go beyond hash-based message authentication.

1. INTRODUCTION

A secret can be easily inserted in the locked and sealed piggy-bank but it cannot be withdrawn without opening the box. In the cryptographic applications of this idea, a signed letter that lists and certifies the contents of the box is sent together with the sealed piggy-bank [1]. We show how the idea of the piggy bank can strengthen the double-lock (DL) cryptography that is behind methods such as the Diffie-Hellmann key exchange [2] and the three-stage quantum cryptography protocol [3].

The basic DL protocol does not address the question of authentication of the two parties and, therefore, this scheme suffers from the man-in-the-middle (MIM) attack. The timing information can be used to detect the MIM attack. The MIM attack is easier to deal with in communications systems where there are no latencies in the transmission as compared to the case of computer networks where the attack can be facilitated by DNS spoofing. In computer networks, the forensics related to an MIM attack include time-to-live (TTL) count of the IP packets, round-trip time between the client and the server which can be estimated by measuring the time from when the client sends its initial TCP SYN packet to when it receives a TCP SYN+ACK from the server. A low number of router hops as well as fast response time will be an indicator of an IP MIM attack.

In this note basic issues associated with the MIM attack are examined, The piggy bank approach provides a means of authentication of the contents in a computer network and thereby it provides protection against such attack.

2. THE BASIC DOUBLE-LOCK SCHEME AND THE MIM ATTACK

To consider the basic DL scheme imagine Alice wishes to send a secret gift, S, to Bob via an untrustworthy courier. We assume that both Alice and Bob have access to unbreakable boxes, locks and keys. Alice places the gift within such a box and sends it to Bob after locking it with her key A. Bob adds his own lock to the box and sends it back to Alice. Alice unlocks her lock and sends it back to Bob who can then unlock it and receive the gift meant for him. The locks of Alice and Bob are labeled A and B.

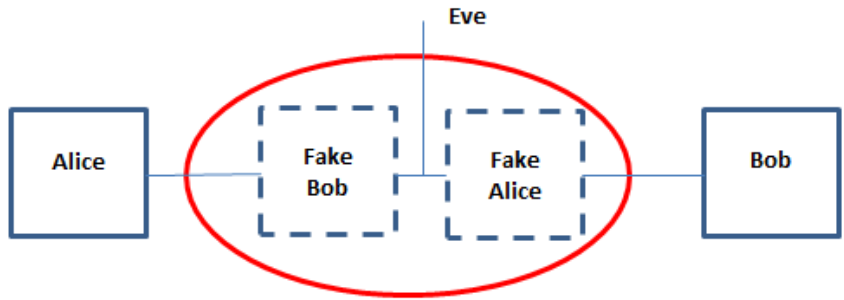

Figure 1. The MIM attack by Eve

In the MIM attack, Eve, in collusion with the courier, interposes herself between Alice and Bob, telling Alice that she is Bob and telling Bob that she is Alice (Figure 1). When she receives the box sent by Alice, she puts her own unbreakable lock, E, on the Box and sends it back to Alice who innocently assumes that it came from Bob. She unlocks her lock and sends the box to Eve who is able to open it and get the gift S. Meanwhile, she has also sent a fake gift, F, to Bob in another box which she locks with her lock and has the courier take it to Bob with the word that it has come from Alice. Bob puts his lock on it, sends it to Eve who unlocks her lock and sends it back to Bob, who now receives F.

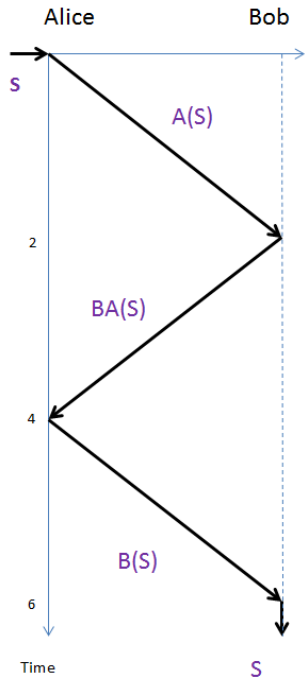

Figure 2. Normal operation of the DL protocol

Since in a single consideration of the application of DL Bob does not get what Alice sent him, there are ways for the two parties to know that the true gift did not arrive at the destination by means of a separate communication. In the case of data, the two parties could compare the hash digests of the sent and the received message.

When we consider the repeated use of the MIM attack on messages (rather than gifts) from Alice to Bob, we see that Eve can send Alice's messages to Bob (and vice versa) just one step behind (Figure 3).

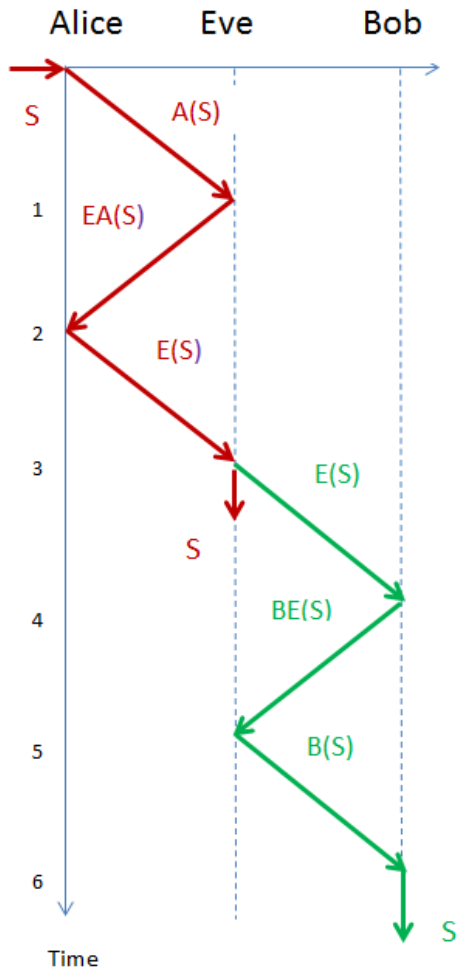

Figure 3. MIM with timing information

Therefore, any signature appended by Alice to the message will be forwarded by Eve to Bob. However, if each of the operations came with spoof-proof time-stamps, the action of Eve will become transparent.

## 3. TIMING INFORMATION

One way to counter the MIM attack is to use the timing information associated with the data, if such information is available. In the normal functioning of the protocol as in Figure 2, the data leaves Alice at time $t_A(1)=0$ and reaches her back at $t_A(2)==4$. Let this time between two consecutive exchanges for Alice be called $T_A$, which has a value of 4. The corresponding times for Bob are $t_B(1)=2$, and $t_B(2)=6$, and thus $T_B= 4$. In the case of MIM attack as in Figure 3, the timings are:

    Alice: 0 and 2
    Bob: 4 and 6

Thus $T_A(MIM) = T_B(MIM) = 2$, whereas $T_A(normal) = T_B(normal) = 4$.

Obviously the difference is due to the fact that the distances between Alice and Eve as well as Eve and Bob are less than between Alice and Bob. Here we assume that Eve is midway between Alice and Bob. If that is not the case then there would be additional asymmetry between the total times for Alice and Bob. For example if Eve is as far away from both Alice and Bob as they are from each other, the total time in the DL protocol would be doubled and this can also be a sign of the attack.

The timing information in the normal case and in the MIM attack is represented in Figure 4.

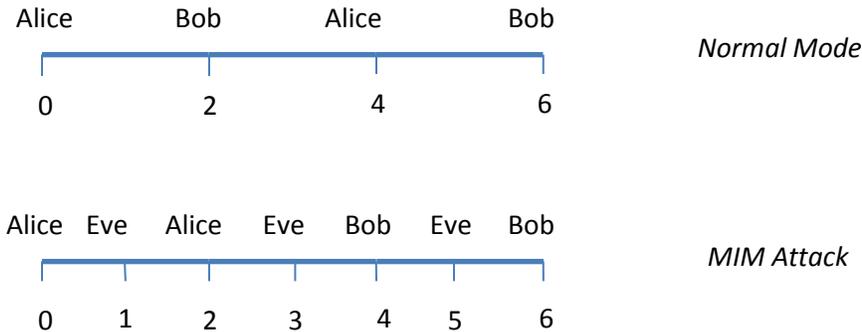

Figure 4. Timing information (for Eve situated in the middle)

The means of the two times for Alice and Bob in the normal mode are 2 and 4, respectively; in other words, the mean time for Bob is twice the mean time for Alice. In the case of the MIM attack, the means are 1 and 5, respectively, or the ratio of the two means has increased to 5. Additionally, the time between the two communications for both Alice and Bob has gone down from the original 4 units to the lower 2 units.

Both these pieces of information can be used to determine if an MIM attack has taken place. The use of the first of these pieces of information requires a comparison between the values for Alice and Bob and, therefore, a secure exchange that is not accessible to Eve. The second piece of information can be processed by each of these parties separately.

The DL protocol requires a handshake protocol prior to the transformations of Figure 2 for authentication so that the timing comparisons can be made. We propose the use of the piggy bank idea in exchange of the relevant timing information.

## 4. THE PIGGY BANK TROPE

In the piggy bank system, the box and the contents authenticate the parties. Bob sends an empty locked, unbreakable piggy bank to Alice who, when she receives it, deposits the secret (money, bills, and jewels) into the box together with the decryption key of a signed letter. In addition, she prepares a letter to be sent separately. The piggy bank and the letter are sent back to Bob. The letter is required to authenticate the contents of the locked piggy bank box. It cannot be in plaintext because the content list itself is a secret. The letter is needed to establish the identity of the person who has sent the secret (that is Alice) and this may carry an additional secret. The letter may be sent within the box if there is certainty that it cannot be taken out by the untrustworthy courier.

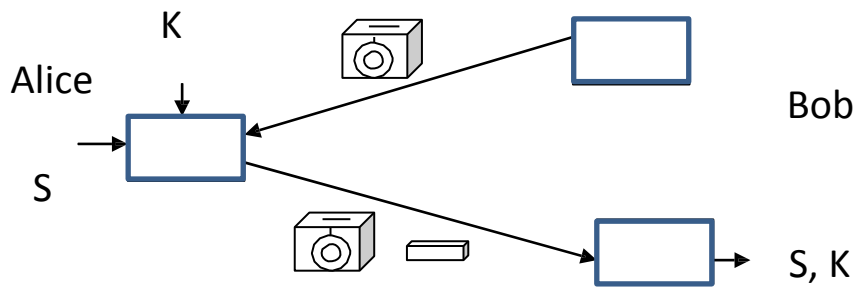

Figure 5. The piggy bank cryptographic trope; the secret letter is represented by

Bob opens the box, obtains the secret, and also reads the signed letter which has further details of the secrets in it. This method is described greater detail in [1] and summarized in Figure 5.

As shown in Figure 5, the box authenticates Bob and Alice is authenticated by the signed letter and this provides a manner of linking the message to the specific communication that was lacking in the DL protocol. We have deliberately not specified the form of the letter for this can be done in a variety of ways depending on the application in mind. In particular, one can imagine a special kind of "letter" in which some identification of the secret is added to the signature of Alice. When Bob obtains the secret, he can compute the identification sequence and then use it to determine the signature of Alice and compare it to the one on file.

In continuing communication, the initial authentication secures the first exchange. This, in turn, can be linked to the security of the second message, and so on. The idea of linkages of messages has been used in the past in a variety of situations. A method of authentication when messages are exchanged is described in [4]. An interesting method was devised to deal with the possibility of dishonest postal employees stealing letters containing currency notes over a hundred years ago in Kashmir in which a unique operation performed on the notes authenticates the contents (for the milieu of that age, see [5]). According to my father, his father would first cut currency notes in half, and send one set of halves in a sealed registered letter through the post office. Only after he had received confirmation that the letter had arrived safely would he send another letter with the missing halves in another registered cover. Once both sets had arrived, the notes would be spliced together and presented to the bank to be replaced (which was required by law to accept damaged or spliced currency notes as long as the serial number was legible). The risk taken by my grandfather was that the dishonest employee would simply destroy the contents of the pilfered cover, but he reasoned that while some employees might be dishonest they were not likely to be malicious.

To go back to the question of the MIM attack on the DL protocol, observe that initially neither Alice nor Bob have side information available for authenticating each other. The idea of using only half the information cannot be used since for Alice and Bob, Eve is invisible and received data (which has only been read and not altered) will pass all tests of authentication.

We now consider a variation of the DL scheme in which the idea of an accompanying coded number provides initial authentication (of course, other schemes of authentication [6] can also be used). And since it is based on the piggy bank trope we call this method Double Signature Double Lock Tropocol (from trope).

## 5. THE DOUBLE-SIGNATURE DOUBLE-LOCK TROPOCOL (DDT)

In the normal operation of DDT, as shown in Figure 6, Alice and Bob share a random number R, distributed to them by an Authenticating Authority at the end of an initial handshake.

Alice sends A(S) together with the hash of R+A(S). After receiving A(S), Bob can confirm that Alice is in possession of R, which authenticates Alice to Bob. In his turn, Bob adds his signature which is the hash digest of R and BA(S) and sends it together with BA(S) back to Alice. Alice can compute her own digest and confirm that what she has received is correct, thereby authenticating Bob. (If Alice's own signature doesn't match Bob's she terminates the communication, and likewise for Bob at the end of the first pass.) This double signature makes it impossible for Eve to mount the MIM attack since she doesn't possess R.

The reason why Alice and Bob compute the hash of R+A(S) as well as R+BA(S) is to prevent replay attack.

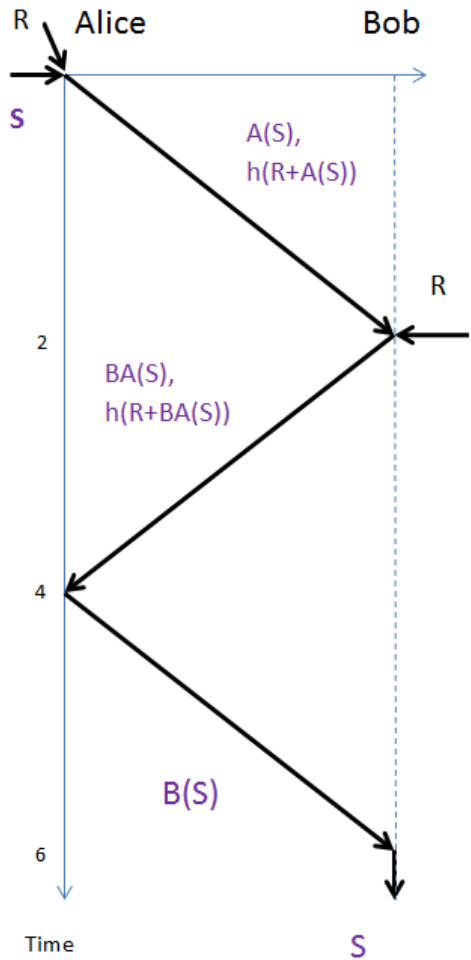

Figure 6. Normal operation of DDT

DDT can be strengthened by the use of additional timing information. Alice and Bob can introduce random delays in their processing of incoming data. Assuming there are no other delays beyond the travel time of the messages, both Alice and Bob can use delays that represent binary random sequences. Furthermore, if the sequences used by Alice and Bob are orthogonal or near-orthogonal (e.g. [7]-[9]) (and this can be assured by the initial handshake), they can each compute the autocorrelation of the delay process to confirm that the data that is being received indeed came from them in an earlier pass.

The double-signature can also be performed implicitly. Alice can send $S_1$+R in the first exchange, $S_2$+$S_1$ in the second exchange and so on. Without the knowledge of R, the reading of the messages will not be useful to Eve although she could mount other cryptographic attacks on the intercepted data. But in an augmented security system where the timing information is also being used, such data will in any way be denied to Eve.

## 6. CONCLUSIONS

The piggy bank idea, which allows one-way encryption of information that can be accessed only by authorized parties, has been employed to authenticate parties to counter man-in-the-middle (MIM) attack. In some versions, this may be considered a special case of security enhancement using hash-based message authentication code; in other versions, such as implicit security, the implementation does not require the use of hashing.